\documentclass[twocolumn,showpacs,preprintnumbers,draftclsnofoot]{revtex4}
\usepackage{amsmath}
\usepackage{amssymb}
\usepackage{graphicx}
\usepackage{dcolumn}
\usepackage{bm}
\usepackage{amssymb}
\usepackage{graphicx}
\usepackage{dcolumn}
\usepackage{bm}
\begin{document}

\title{ Wavy optical grating: wideband reflector and Fabry-Perot BICs }
\author{Ma Luo\footnote{Corresponding author:swym231@163.com}, Feng Wu }
\affiliation{School of Optoelectronic Engineering, Guangdong Polytechnic Normal University, Guangzhou 510665, China}

\begin{abstract}

In this study, we theoretically and numerically investigate the resonant modes and reflectance of an optical grating consisting of a wavy dielectric slab by applying the spectral element method. The presence of the wavy shape transforms the waveguide modes into leaky resonant modes. A few resonant modes with specific longitudinal wave number have infinitely large Q factor, while the other resonant modes have finite Q factor. For the leaky resonant mode with zero longitudinal wave number, the Q factor is inversely proportional to the amplitude of the wavy shape. An array of multiple low-Q wavy gratings has a high reflectance in a large bandwidth. A double-layer wavy grating forms a Fabry-Perot cavity, which hosts Fabry-Perot bound states in the continuum (BICs) at the resonant frequency. The Q-factor of the Fabry-Perot cavity can be tuned by adjusting the distance between the two wavy slabs. The wavy shape could be generated by a vibrational wave in a flat dielectric slab so that the BICs mode and wideband reflectance could be controlled on-demand.

\end{abstract}

\maketitle

\section{Introduction}

Bound states in the continuum (BICs) have been intensively studied in recent years because of light confinement in the absence of an energy gap \cite{Sadreev21,Marinica08,Bulgakov08,Plotnik11,BoZhen14,YiYang14} and the potential functionality for various types of photonic devices, such as lasing and sensing devices. The electromagnetic field of the BICs is confined to the systems because of the symmetry mismatch between the bounded states and radiative states \cite{Koshelev18}, so that the energy levels of the localized BICs are embedded into the continuous energy spectrum of the delocalized radiative modes \cite{fengwu19}. Theoretically, BICs have an infinitely large Q factor and do not couple with the incident travelling wave; therefore, the application of BICs is not feasible. Nevertheless, quasi-BICs mode has ultra-high Q factor, and has coupling with the incident travelling wave, so that multiple applications based on the quasi-BICs mode have been proposed, such as lasers \cite{Hirose14,Kodigala17,Hwang21}, sensors \cite{Romano18,XChen20,Srivastava19,Maksimov16}, light absorption \cite{Xwang20,SXiao20,SCao20,TSang21}, and enhancement of harmonic generation \cite{ZLiu19,TNing20,Koshelev20,JWang20,ZHuang21}.

Among multiple designs of optical systems with quasi-BICs, the compound optical grating waveguide system enables experimental observation of quasi-BICs owing to its simple structure \cite{fengwu19,fengwu21}. Quasi-BICs appear when the reciprocal wave number of the grating structure is equal to the wave number of the waveguide mode. Theoretical studies have found that quasi-BICs can significantly enhance optical $Goos$-$H\ddot{a}nchen$ shifts \cite{fengwu21}. However, the structures of these systems are relatively complicated to implement experimentally. Specifically, the strict periodicity of the grating layer and the accuracy of the small shift in the sublattice elements of the grating are difficult to obtain.

In this paper, we propose a simple structure consisting of a monolayer of an air-bridge wavy dielectric slab that supports leaky resonant modes. The wavy shape of the dielectric slab can be obtained using a piezoelectric mechanical oscillator to launch a transverse oscillating standing wave at a flat dielectric slab \cite{Teufel08,KwanLee10,Aspelmeyer14}. The period and amplitude of the standing wave control the resonant frequency and Q-factor of the leaky resonant mode. At the resonant frequency, the reflectance was close to unity. The spectral element method (SEM) is applied to numerically calculate the reflectance and electromagnetic field patterns \cite{SEM1,SEM2,SEM3,SEM4,SEM5,SEM6}.

On the other hand, various applications require mirrors with high reflectance in large bandwidths. The most widely used optical system for obtaining high reflectance is the distributed Bragg reflector (DBR) \cite{Fink98,Anguiano14,Bikbaev17}, which consists of multiple bilayers of quarter-wavelength dielectric slabs with different refractive indices. The interference between the reflections of each interface is constructive, and thus, the reflection is enhanced. However, to obtain a high reflectance that is close to one, the number of bilayer dielectric slabs must be as large as 30, which in turn increases the size of the optical devices. The bandwidth of the high reflectance can be increased by increasing the contrast of the refractive index of the neighboring layers; however, the necessary number of bilayers to obtain a large reflectance is still no less than 10. Another scheme to obtain nearly one reflectance is to utilize quasi-BICs. At the resonant frequency of the quasi-BICs mode, the reflectance was approximately one. However, the bandwidth of high reflectance in systems with quasi-BICs is narrow.

In this study, we found that an array of multiple wavy gratings with low-Q leaky resonant modes has high reflectance over a large bandwidth. Only two to four layers of the low-Q wavy grating can obtain a high reflectance in a large bandwidth. The bandwidth of the high reflectance is tuned by the relative shift between the wavy shapes of the neighboring layers along the periodic direction. The total thickness of the entire system can be as small as its wavelength. In a double-layer wavy grating, the Fabry-Perot BICs \cite{Marinica08,Bulgakov18,Hemmati19,HXu20,fengwu22} appear owing to the coupling between the leaky resonant modes of the two wavy layers. As the distance between the two layers varied, the Q factor of the Fabry-Perot cavity was tuned. The proposed system can be applied to Q-switching devices with large bandwidths.

The remainder of this paper is organized as follows. II, the structure of the proposed system and the numerical simulation method are described. In Sec. III, the numerical results of the monolayer wavy grating with high-Q leaky resonant mode, the multiple layer wavy grating with large bandwidth high reflectance, and the double layer wavy grating with Fabry-Perot BICs are summarized. In Sec. IV, conclusions are presented.

\section{Structure and Theoretical model}

\begin{figure}[tbp]
\scalebox{0.6}{\includegraphics{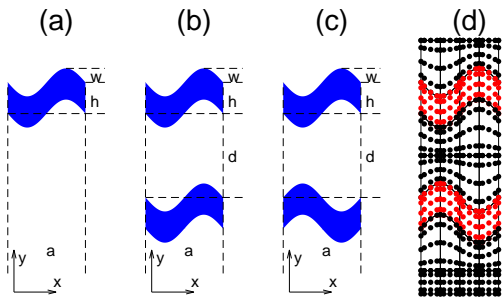}}
\caption{ (a-c) Structure of the wavy grating (array) in one unit cell. The structure is periodic along x axis. The thickness of each dielectric slab is $h$, the amplitude of the wavy shape is $w$, the period of the wavy shape is $a$, the distance between two neighboring wavy gratings is $d$. $\phi_{0}$ equates to $0$ and $\pi$ in (b) and (d), respectively. (d) Discritization of the structure by curvilinear quadrilateral elements and distribution of the nodal points. }
\label{figure_0}
\end{figure}

The structure of the proposed system is shown in Fig. \ref{figure_0}. In Fig. \ref{figure_0}(a), a monolayer wavy grating is plotted with the structure parameters being indicated. The refractive index of the dielectric slab was $n=2$, and the refractive index of the background was one. The wavy shape, that is, the function of the change in the y coordinate of the top and bottom boundaries of the dielectric slab, is $w\sin(2\pi x/a)$, where $a$ is the period of the wavy shape and $w$ is the amplitude of the wavy shape. $a$ was assumed to be 333 nm. $h$ is the thickness of the dielectric slab, assumed to be 134 nm. The structure of the double-layer wavy grating is illustrated in Fig. \ref{figure_0}(b,c). The distance between the two layers is $d$ and the wavy shape of the second layer is $w\sin(2\pi x/a+\phi_{0})$. The wavy shapes of the two layers have a relative phase shift $\phi_{0}$, with $\phi_{0}=0$ in (b) and $\phi_{0}=\pi$ in (c). In our numerical simulation, we considered three situations: $\phi_{0}=0$, $\pi/2$, and $\pi$. In the presence of more layers, the odd (even) layers had the same wavy shape as the first (second) layer. When $\phi_{0}=\pi$, the structure is designated as a multiple-layer staggered wavy grating.

Transverse electric (TE) mode polarization at normal incidence is considered so that the optical field is governed by the two-dimensional Helmholtz equation. The spectral element method (SEM) was applied to discretize the equation. The structure shown in Fig. \ref{figure_0}(a-c) is split into curvilinear quadrilateral elements that are conformal to each other and cover the whole region. The basis functions in each element are constructed by Gauss-Lobatto-Legendre(GLL) polynomials in a reference element and are mapped to each element by covariant mapping. The distribution of nodal points in real space is shown in Fig. \ref{figure_0}(d). A periodic boundary condition was applied at the left and right boundaries. The radiation boundary condition given by the spectral integral method (SIM) was applied at the top and bottom boundaries. The total-field/scattering-field method was applied to simulate the incidence of the plane waves. The SEM-SIM hybrid method has been proven to offer high accuracy and efficiency in the simulation of optical scattering in grating structures \cite{SEM3}. The reflectance and field pattern can be obtained by postprocessing the numerical results. On the other hand, in the absence of an incident field, the resonant mode can be obtained by solving the nonlinear eigenvalue problem given by the weak form of the SEM-SIM hybrid method.

\section{Numerical result}

\subsection{resonant modes of monolayer wavy grating}

\begin{figure*}[tbp]
\scalebox{0.58}{\includegraphics{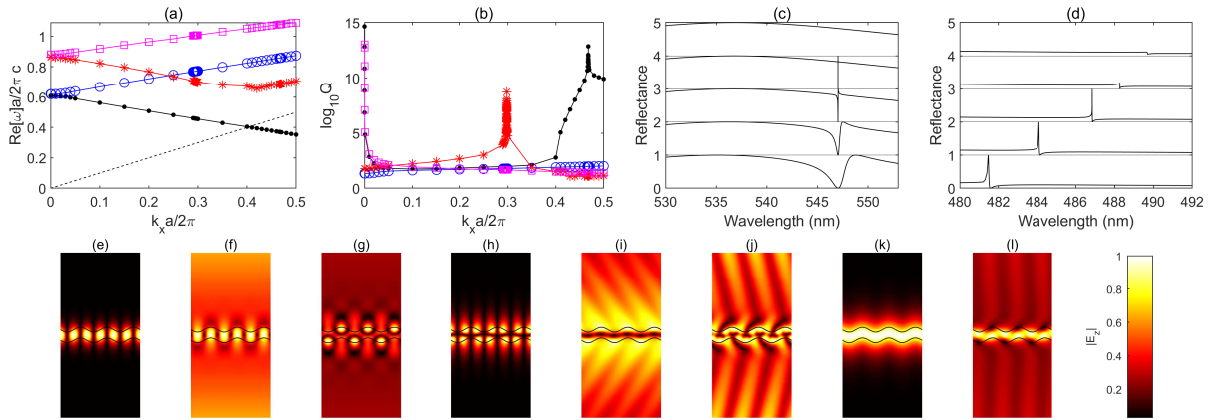}}
\caption{ (a) The resonant frequency of the monolayer wavy grating with $w=30$ nm versus the wave vector $k_{x}$ for the lowest four resonant modes. The first to the fourth resonant modes are plotted as black(dotted), blue(circled), red(stared), and magenta(squared) lines, respectively. The dashed line is the light cone. (b) The Q factor of the corresponding resonant mode in (a) is plotted. (c) TE reflectance spectra of the monolayer wavy grating with $k_{x}a/2\pi$ being 0.01, 0.005, 0.001, 0.0001, and 0 for the five lines from bottom to top. (d) TE reflectance spectra of the monolayer wavy grating with $k_{x}a/2\pi$ being 0.308, 0.312, 0.316, 0.318, and 0.32 for the five lines from bottom to top. (e-h) The electric field distributions ($|E_{z}|$) of the first to fourth resonant modes at $k_{0}=0$. (i,j) The electric field distributions ($|E_{z}|$) of the third and fourth resonant modes at $k_{x}a/2\pi=0.316$. (k,l) The electric field distributions ($|E_{z}|$) of the first and second resonant modes at $k_{x}a/2\pi=0.4679$.  }
\label{figure_mode}
\end{figure*}

The dispersion of the waveguide mode of the flat dielectric waveguide in air background is given by the relation $\tan(h\sqrt{nk_{0}^{2}-k_{x}^{2}}/2)=\sqrt{k_{x}^{2}-k_{0}^{2}}/\sqrt{nk_{0}^{2}-k_{x}^{2}}$ for the even-mode, and $\tan(h\sqrt{nk_{0}^{2}-k_{x}^{2}}/2)=-\sqrt{nk_{0}^{2}-k_{x}^{2}}/\sqrt{k_{x}^{2}-k_{0}^{2}}$ for the odd-mode, with $k_{0}=2\pi/\lambda$, $\lambda$ being the wavelength, and $k_{x}$ being the wave number along the waveguide. If periodic perturbation of the structure with period $a$ is imposed on the waveguide, the dispersion of the resonant guided modes is modified. The value of $k_{x}$ is restricted to $[0,2\pi/a]$ because the optical field in each unit cell satisfies the periodic boundary condition, with the Bloch phase being $e^{ik_{x}a}$. When the perturbation is small, the resonant guided modes with $k_{x}$ approximately consist of standing waves, which are the superposition of two counter-propagating waveguide modes with wave number $k_{x}\pm\frac{2\pi}{a}$. Because the wavy structure shown in Fig. \ref{figure_0}(a) have inversion symmetric, for the cases with $k_{x}=0$, the node of the standing wave could be at the node or antinode of the wavy shape. Thus, two resonant modes at $k_{x}=0$ were constructed using one dispersive band of the waveguide mode. As $k_{x}\ne0$, the two resonant modes evolve into two dispersive bands of resonant modes. We studied the resonant modes of a specific monolayer wavy grating with $d=30$ nm by applying the SEM/SIM hybrid method; the numerical results are summarized in Fig. \ref{figure_mode}. The real part of the resonant frequency $\omega$ of the four bands of the resonant modes versus $k_{x}$ is plotted in Fig. \ref{figure_mode}(a). The first and second bands of the resonant modes, which consist of the standing wave of the first waveguide even-mode, are plotted as black (dots) and blue (circles) lines, respectively. The third and fourth bands of the resonant modes, which consisted of a standing wave of the first waveguide odd-mode, are plotted as the red (start) and magenta (square) lines, respectively. The Q factor of each resonant mode, which is defined as $Q=Re[\omega]/Im[\omega]$ with $Re[\omega]$ and $Im[\omega]$ being the real and imaginary parts of $\omega$, are plotted in Fig. \ref{figure_mode}(b).

Three BICs with infinite Q factors were found in the band structure. At $k_{x}=0$, the Q factor of two resonant modes in the first and the fourth bands is infinitely large, as shown by the black(dotted) and magenta (square) lines at $k_{x}=0$ in Fig. \ref{figure_mode}(b). In each unit cell, the field pattern of the electric field $|E_{z}|$ has two antinodes along the longitudinal direction ($\hat{x}$ direction). The antinodes of the field pattern of the two resonant modes with infinitely large Q factor are at the node of the wavy shape, as shown by the field pattern in Fig. \ref{figure_mode}(e) and (h). Thus, the two antinodes in one unit cell align along a strain line in $\hat{x}$ direction such that the energy flows straight forward without loss. Consequently, the Q-factor is infinitely large. For the other two resonant modes at $k_{x}=0$, the antinode of the field pattern is at the antinode of the wavy shape such that the two antinodes of the field pattern in one unit cell have a staggered transversal location, as shown in Fig. \ref{figure_mode}(f) and (g). As a result, energy flows along the zigzag lines in $\hat{x}$ direction. As the direction of the energy flow changes, part of the energy is lost, so that the Q factor is finite. The resonant modes with infinite and finite Q factors were BICs and leaky resonant modes, respectively. As $k_{x}$ increases slightly, the location of the antinode of the field pattern moves forward along the wavy slab. Owing to the wavy shape of the dielectric slab, the two antinodes of the field pattern in one unit cell move along the opposite transverse direction. Thus, the antinodes of the non-leaky resonant modes become non-aligned along a straight line, resulting in energy loss. As a result, the Q factor decreases sharply, and the BICs become quasi-BICs. The TE reflectance versus wavelength $\lambda$ of the incident plane wave with a fixed $k_{x}$ is plotted in Fig. \ref{figure_mode}(c). For the five lines from the bottom to the top of the figure, $k_{x}$ is fixed at a different value. Note that the incident angle of the plane wave is not fixed for each line but depends on $\lambda$ as $\theta_{in}=\sin^{-1}[k_{x}\lambda/(2\pi)]$. At the resonant wavelength of the first leaky resonant mode, the reflectance was approximately one. At the resonant wavelength of the first non-leaky resonant mode, the reflectance has no peak at $k_{x}=0$, because the incident plane wave cannot excite the BIC. As $k_{x}$ becomes nonzero, a sharp peak with a Fano shape appears because the incident plane wave excites quasi-BIC. As $k_{x}$ moves further away from zero, the resonant peak becomes wider because the Q factor of the quasi-BIC becomes smaller. A similar phenomenon occurs when the $\lambda$ of the incident wave is near the resonant wavelength of the second non-leaky resonant mode.

The third BIC appeared in the third band of the resonant modes at $k_{x}a/2\pi=0.316$, as shown in Fig. \ref{figure_mode}(b). The resonant frequency of the resonant mode with an infinite Q factor is above the light cone such that the non-leaky resonant mode coexists with the radiating continuum. The TE reflectance versus wavelength $\lambda$ of the incident plane wave with a fixed $k_{x}$ is plotted in Fig. \ref{figure_mode}(d). For the five lines from the bottom to the top of the figure, $k_{x}a/2\pi$ are fixed at five values near the BIC wave number, that is, $0.316$. A Fano-shaped resonant peak with a finite width appears at the resonant wavelength of the leaky resonant mode. As $k_{x}a/2\pi$ moves further away from $0.316$, the width of the peak increases. As $k_{x}a/2\pi<0.316$, the reflectance at the top of the peak is equal to one, so the corresponding leaky resonant mode is quasi-BIC. In contrast, when $k_{x}a/2\pi>0.316$, the reflectance at the top of the peak is significantly smaller than one, so that the corresponding leaky resonant mode is not quasi-BIC. With $k_{x}a/2\pi=0.316$, the field patterns of the resonant modes in the third and fourth bands are plotted in Fig. \ref{figure_mode}(i) and (j), which are BIC and leaky resonant mode, respectively. For the BIC in Fig. \ref{figure_mode}(i), the maximum of the field pattern concentrate at the antinode of the wavy shape at the two edges of the dielectric slab. The maximum of the field pattern aligns along a straight line so that the energy flow is lossless. The BIC localization is weak because a large portion of the field pattern is distributed outside the dielectric slab. For the leaky resonant mode shown in Fig. \ref{figure_mode}(j), the maximum of the field pattern concentrate at the node of the wavy shape at the two edges of the dielectric slab, and then is reflected to the opposite edges. Thus, the energy flows along a zigzag line with loss. The lost energy radiates outside the dielectric slab as an obliquely propagating plane wave.

In addition to the three BICs, a resonant wavy mode with an infinite Q factor appeared under the light cone. As $k_{x}a/2\pi>0.4$, the first band of the resonant modes is under the light cone, and thus, the corresponding resonant modes have a large Q factor. As $k_{x}a/2\pi$ reaches a specific value of $0.4679$, the Q-factor becomes infinitely large. The field patterns of the resonant modes of the first and second bands at $k_{x}a/2\pi=0.4679$ are shown in Fig. \ref{figure_mode}(k) and (l), respectively. The resonant modes shown in Fig. \ref{figure_mode}(k) has infinitely large Q factor. The field pattern is uniform along the localized axis of the wavy dielectric slab, which is the same as mapping the field pattern of the traveling waveguide mode (not a standing wave) of a flat dielectric slab into a wavy shape. The field travels along a wavy path that exactly matches the wavy shape along the axis of the dielectric slab so that the energy flows without loss. Thus, the resonant mode is designated as a non-leaky resonant wavy mode. The resonant modes shown in Fig. \ref{figure_mode}(l) has small Q factor, because the mode is above the light cone. The field pattern is concentrated in a zigzag path within the dielectric slab. The width of the field pattern at the path was narrower than that of the dielectric slab. The zigzag shape of the path and wavy shape of the dielectric slab have a 90$^{o}$ phase shift along $\hat{x}$ direction. As the path reached one edge of the dielectric slab, the optical field was reflected so that the path turned to the opposite edge. The internal reflection has a significant loss owing to the curved shape of the edge, so that the energy flow is lost. The resonant mode was designated as the leaky resonant zigzag mode.

\begin{figure}[tbp]
\scalebox{0.6}{\includegraphics{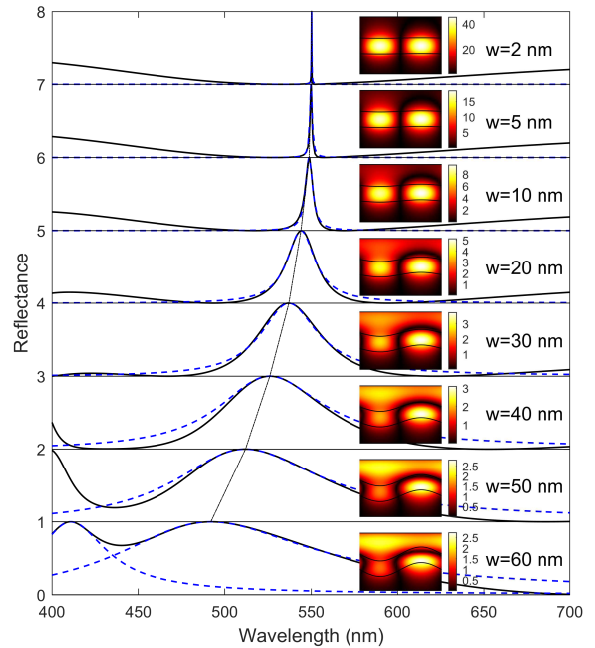}}
\caption{ TE reflectance spectra around $\lambda=550$ nm of the monolayer wavy grating for different values of $w$ under normal incidence. The dashed-dotted line connect the resonant peaks of varying $w$. The blue dashed lines are the fitting curve of the Lorentzian line shape. The insets represent the electric field distributions ($|E_{z}|$) at the corresponding reflectance peaks, with the amplitude of the incident plane being one. }
\label{figure_1}
\end{figure}

\begin{table}[htpb]
\centering
\caption{Fitting parameters of the resonant frequency, radiative loss, and the coupling strength, which are normalized by the frequency $\omega_{r}=2\pi c_{0}/\lambda_{r}$ with $\lambda_{r}=$550 nm. }
\label{table1}
\begin{tabular}{|c|c|c|c|}
\hline
$w$ (nm) & $\omega_{0}/\omega_{r}$ & $\gamma_{0}/\omega_{r}$ & $-\kappa_{0}/\omega_{r}$ \\
\hline
2 & 0.9998 & 2.999$\times10^{-4}$ & 3.464$\times10^{-3}$ \\
\hline
5  & 1.000 & 1.501$\times10^{-3}$ & 1.980$\times10^{-3}$ \\
\hline
10 & 1.002 & 4.511$\times10^{-3}$ & 7.792$\times10^{-4}$ \\
\hline
20 & 1.011 & 1.820$\times10^{-2}$ & 1.207$\times10^{-4}$ \\
\hline
30 & 1.024 & 3.790$\times10^{-2}$ & 1.869$\times10^{-5}$ \\
\hline
40 & 1.046 & 7.319$\times10^{-2}$ & 2.896$\times10^{-6}$ \\
\hline
50 & 1.074 & 1.074$\times10^{-1}$ & 4.485$\times10^{-7}$ \\
\hline
60 & 1.118 & 1.565$\times10^{-1}$ & 6.948$\times10^{-8}$ \\
\hline
\end{tabular}
\end{table}

By varying $d$ for the wavy shape of the grating, the dispersion of the four bands of the resonant modes was changed. In the remainder of this article, we investigate the scattering of normally incident plane waves by wavy grating, that is, scattering with $k_{x}=0$. As the two BICs at $k_{x}=0$ are symmetric mismatches with the normally incident plane wave, they are not excited. In contrast, the normally incident plane wave excites the two leaky resonant modes at $k_{x}=0$. The numerical results of the TE reflectance versus the wavelength of the incident field for varying $d$ are plotted in Fig. \ref{figure_1}. The numerical results show that as $w$ approaches zero, the wavelength of the leaky resonant mode approaches 550 nm, which corresponds to the first even waveguide mode of the flat dielectric slab at $k_{x}=2\pi/a$. The reflectance spectra can be modeled using the temporal coupled mode theory (TCMT) \cite{SFan03,fengwu22}, which provides a Fano line shape. At the resonant wavelength, the direct reflection rate is near zero. Thus, the Fano line shape can be approximated by the Lorentzian line shape as $R=\gamma_{0}^{2}/[(\omega-\omega_{0})^2+\gamma_{0}^{2}]$, where $\omega$ is the frequency of the incident wave, $\omega_{0}$ is the resonant frequency of the leaky resonant mode, and $\gamma_{0}$ is the radiative loss. The parameters $\omega_{0}$ and $\gamma_{0}$ for the wavy grating with varying $w$ can be obtained by fitting the numerical result, which are listed in Table \ref{table1}. The fitted values of $\omega_{0}$ and $\gamma_{0}$ are the same as the real and imaginary parts of the eigenvalue of the resonant frequency of the leaky resonant mode, respectively. When $w=2$ nm, the Q factor is as large as $3.3\times10^{3}$. As $w$ increases, the Q factor decreases, and the resonant wavelength, which is $\lambda_{0}=2\pi c_{0}/\omega_{0}$ with $c_{0}$ being the speed of light in vacuum, decreases (or $\omega_{0}$ increases). The field pattern of the out-of-plane electric field $|E_{z}|$ at the resonant wavelength shows that as $w$ decreases, the localized electromagnetic field is strongly excited at the resonant wavelength. When $w=2$ nm, the incident plane wave excited the leaky resonant mode, so that the amplitude of the electric field inside the dielectric slab was 40 times larger than that of the incident plane wave. The field pattern inside the dielectric slab is nearly the same as that of the standing wave of the opposite travelling waveguide mode of the flat dielectric slab with the same thickness.

\subsection{Wideband reflectance of multiple layer wavy grating}

\begin{figure}[tbp]
\scalebox{0.6}{\includegraphics{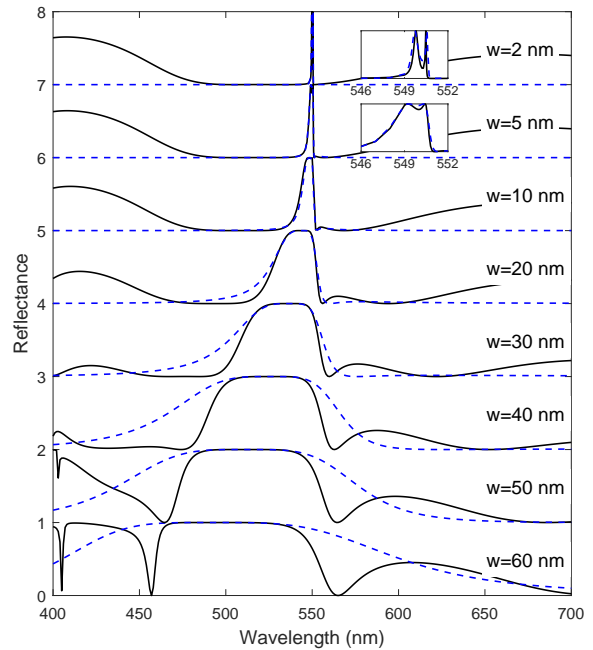}}
\caption{ TE reflectance spectra around $\lambda=550$ nm of the double layer wavy grating for different values of $w$ under normal incidence. $d=360$ nm. The insets zoom in the peak for better visualization. }
\label{figure_2}
\end{figure}

If two layers of wavy gratings are placed together in parallel with the distance between them being $d=360$ nm, the reflectance spectrum is determined by the coupling between the leaky resonant modes of the two wavy gratings. The numerical results for the double-layer wavy grating are shown in Fig. \ref{figure_2}. In this case, $\phi_{0}=\pi$. Two types of coupling mechanisms coexist: near- and far-field coupling. The near-field coupling is caused by the overlap between the mode patterns of the two resonators. The strength of near-field coupling is denoted by $\kappa_{0}$. For a double-layer wavy grating, $\kappa_{0}$ is exponentially dependent on the distance between the two layers. Because the two layers of the wavy grating are relatively shifted for a half-period, the sign of $\kappa_{0}$ is negative. The far-field coupling can be decomposed into three steps: one wavy grating radiates a propagating wave due to radiative loss with amplitude $\sqrt{\gamma_{0}}\tilde{a}_{i}$, where $\tilde{a}_{i}$ is the mode amplitude at the $i^{th}$ layer; the propagating wave travels distance $d$ with wave number $k_{0}=\omega_{0}/c_{0}$ and reaches another wavy grating with phase factor $e^{ik_{0}d}$, and the incidence of the propagating wave couples into the leaky resonant mode of the other wavy grating with a coupling strength of $\sqrt{\gamma_{0}}$. Thus, the far-field coupling strength between the two wavy gratings was $\gamma_{0}e^{ik_{0}d}$. By applying TCMT, the line shape of the reflectance spectrum is given by the solution of the coupling mode equations \cite{fengwu22,SFan03}. The coupling strength $\kappa_{0}$ can be obtained by fitting the line shape, as summarized in Table \ref{table1}. On the other hand, $\kappa_{0}$ can also be obtained by calculating the overlapping integral between the field pattern of the resonant mode of one layer and the dielectric contrast of another layer. The fitting-line shapes are plotted as dashed blue lines in Fig. \ref{figure_2}, which match with the numerical results of the reflectance spectrum near to the resonant wavelength of individual layer $\lambda_{0}$. Far from $\lambda_{0}$, the difference between the TCMT line shape and the numerical result is due to the extra coupling with the other leaky resonant modes of the individual layer.

When $w$ is small, the Q factor of the leaky resonant mode of the individual wavy gratings is large. The coupling strength $\kappa_{0}$ is larger than the radiative loss $\gamma_{0}$, as shown in Table \ref{table1}, such that the near-field coupling dominates the scattering process. When the two wavy gratings couple, the leaky resonant modes mix with each other. The hybridization of the leaky resonant modes splits the resonant peak into two peaks, as shown by the insets in Fig. \ref{figure_2}. The two peaks have a highly asymmetric shape. The peaks at smaller (larger) wavelengths had larger (smaller) bandwidths. For comparison, for other systems with $\phi_{0}=0$, the sign of the coupling strength $\kappa_{0}$ is positive, such that the peaks at smaller (larger) wavelengths have smaller (larger) bandwidths. As $w$ increases, the Q factor of the leaky resonant mode of the individual wavy grating decreases, and thus the radiative loss $\gamma_{0}$ increases. Meanwhile, the coupling strength $\kappa_{0}$ decreases because the overlap between the mode pattern of one wavy grating layer and the dielectric contrast of the other wavy grating layer decreases. Consequently, far-field coupling becomes more important than near-field coupling. For the extreme case where $\kappa_{0}\approx0$, by neglecting $\kappa_{0}$ in the TCMT equations, the reflectance given by the TCMT can be expressed as
\begin{eqnarray}
&&R(\omega)=1
\\&&-\frac{\eta^{4}}{1+[1+\eta^{2}]^{2}-(1-\eta^{2})2\cos(2k_{0}d)-4\eta\sin(2k_{0}d)},\nonumber
\end{eqnarray}
where $\eta(\omega)=(\omega_{0}-\omega)/\gamma_{0}$. When $\omega=\omega_{0}$, $R(\omega)$ is equal to one, which is the maximum value. For the given parameters $\omega_{0}$ and $\gamma_{0}$, the function of $R(\omega)$ depends only on the value of $2k_{0}d$. When the condition $2k_{0}d=(1+2N)\pi$ is satisfied (with $N$ being an integer), the function $(1-\eta^{2})2\cos(2k_{0}d)+4\eta\sin(2k_{0}d)$ is minimized over a wide range of $\eta$ near zero, so that $R(\omega)$ is maximized in a wide range of $\omega$ near $\omega_{0}$. For a double layer with $w=60$ nm and an incident wavelength near the resonant wavelength at $\lambda_{0}=492$ nm, assuming $N=1$, the optimized distance between the two layers is $369$ nm. Because the surface of each layer has a wavy shape, the effective distance between the two layers is slightly greater than $d$. Thus, we used $d=360$ nm as an example. The reflectance at wavelengths ranging between 478 and 508 nm was nearly one. The total thickness of the double-layer wavy grating is $d+2h+2w=748$ nm.

\begin{figure}[tbp]
\scalebox{0.6}{\includegraphics{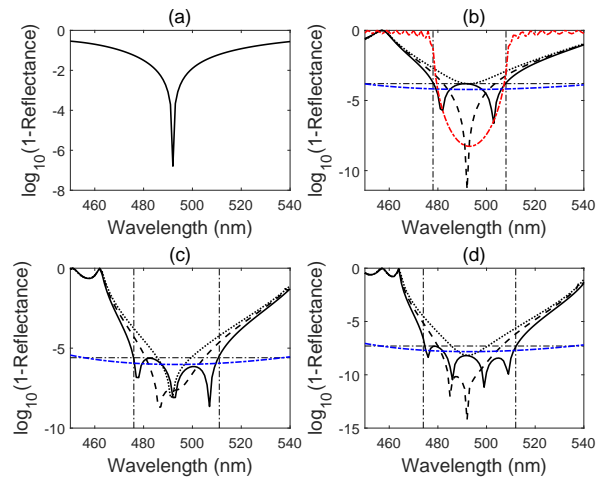}}
\caption{ Base-10 logarithm of one-minus-reflectance of the wavy grating with $w=60$ nm, and with (a) monolayer, (b) double-layers, (c) three layers, (d) four layers. In (b-d), the multiple layers wavy grating with $\phi_{0}$ equating to $\pi$, $\pi/2$, and $0$ are plotted as solid, dashed and dotted lines, respectively. Another structure parameter is $d=360$ nm. The reflectance of the distributed Bragg reflections (DBRs) consisted of 8, 11, and 14 bilayer of quarter-wavelength dielectric slabs with refractive indices of the two slabs being 2 and 1 are plotted in (b), (c) and (d) as blue (dashed-dotted) line, respectively. The DBR consisted of 100 bilayer of quarter-wavelength dielectric slabs with refractive indices of the two slabs being 2 and 1.8166 is plotted in (b) as red (dashed-dotted) line. The two vertical black (dashed-dotted) lines in (b), (c), and (d) mark the range of wavelength, within which the one-minus-reflectance of the double-layer wavy grating with $\phi_{0}=\pi$ is smaller than the value marked by the horizontal black (dashed-dotted) line at $10^{-3.8}$, $10^{-5.6}$, and $10^{-7.3}$, respectively. }
\label{figure_3}
\end{figure}

We further investigated the impact of the relative shift of the wavy shape between the odd and even layer(s) (i.e., $\phi_{0}$) on the reflectance. As a reference, the base-10 logarithm of one minus the reflectance of the monolayer wavy grating is plotted in Fig. \ref{figure_3}(a). A single peak at the resonant wavelength of the leaky resonant mode appeared at a wavelength of 492 nm. In a double-layer wavy grating, when $\phi_{0}=\pi$, the resonant peak splits into two peaks, as shown in Fig. \ref{figure_3}(b). The value of one minus reflectance is smaller than $10^{-3.8}$ in the wavelength range between 478 and 508 nm, as indicated by the vertical dashed-dotted lines in Fig. \ref{figure_3}(b). If $\phi_{0}$ is changed to $\pi/2$, the resonant peak is not split into two peaks. Although the value of one minus reflectance at the resonant wavelength decreases, the bandwidth of the wavelength range in which the value of one minus reflectance is smaller than $10^{-3.8}$ decreases. As $\phi_{0}$ is changed to $0$, the maximum value of the reflectance at the resonant peak is $1-10^{-3.8}$, such that the feature of wide-bandwidth highly reflective disappears. A comparison of the performance of the two distributed Bragg reflections (DBRs) is shown in Fig. \ref{figure_3}(b). For a DBR consisting of eight (100) bilayers of quarter-wavelength dielectric slabs with refractive indices of the two slabs being 2 and (1) 1.8166, the reflectance is plotted as a blue (red) dashed-dotted line. For the DBR with refractive indices of the two slabs being 2 and 1, at least eight bilayers of quarter-wavelength dielectric slabs are required to obtain a reflectance as large as $1-10^{-3.8}$. Consequently, the thickness of the DBR was 1355.8 nm. The DBR with refractive indices of the two slabs being 2 and 1.8166 has the same bandwidth at reflectance $1-10^{-3.8}$ as the double-layer wavy grating with $\phi_{0}=\pi$, while the thickness is 12879 nm. The performances of the two DBRs are similar to that of the double-layer wavy grating, but the thicknesses of both DBRs are larger than those of the double-layer wavy grating.

For a multiple-layer wavy grating with three and four layers, the systems with $\phi_{0}=\pi$ exhibit better performance than the other systems with $\phi_{0}\ne\pi$, as shown in Fig. \ref{figure_3}(c) and (d), respectively. As the number of layers is increased to three (four), the value of one minus the reflectance is $10^{-5.6}$ ($10^{-7.3}$) in the wavelength range between 476 nm and 511 nm (474 nm and 512 nm), as shown in Fig \ref{figure_3}(c)[(d)]. As a result, increasing the number of layers sharply increases the bandwidth of high reflectance. Meanwhile, the thicknesses of the three- and four-layer wavy gratings were increased to $2d+3h+2w=1242$ nm and $3d+4h+2w=1736$ nm, respectively. In order to obtain the same reflectance in a DBR with refractive indices of the two slabs being 2 and 1 as that in the three- and four-layer wavy grating, 11 and 14 bilayers of quarter-wavelength dielectric slabs are required, so that the thickness of the DBRs is 1910.4 nm and 2465 nm, respectively. Consequently, the thickness of the DBRs is larger than that of the multiple-layer wavy grating.

\subsection{Fabry-Perot BICs of double-layer wavy grating}

\begin{figure}[tbp]
\scalebox{0.8}{\includegraphics{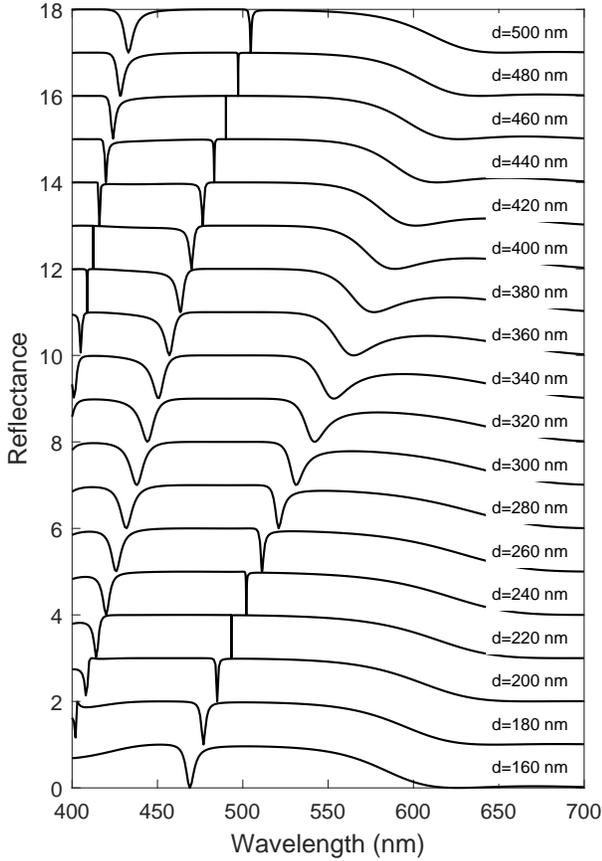}}
\caption{ TE reflectance spectra around $\lambda=550$ nm of the double layer wavy grating for different values of $d$ under normal incidence. $w=60$ nm. }
\label{figure_4}
\end{figure}

In this subsection, the Fabry-Perot BICs of a double-layer wavy grating are discussed. The amplitude of the wavy shape is fixed at $w=60$ nm. The double-layer wavy grating with $\phi_{0}=\pi$ has mirror reflection symmetry about the plane between the two layers. At the resonant frequency of the leaky resonant mode $\omega_{0}$ of the individual wavy grating, each monolayer is an ideal reflector. In the double-layer wavy grating, the electromagnetic wave is reflected back and forth within the spacing between the two layers, so that the system becomes a Fabry-Perot cavity. The resonant mode of the Fabry-Perot cavity can be modeled by TCMT or numerically calculated using the SEM/SIM hybrid method. If the distance between the two reflectors satisfies the condition $k_{0}d=m\pi$ with $m$ being an integer, the interference between the leaky resonant modes of the two layers forms the Fabry-Perot BIC with a resonant frequency of $\omega_{0}+\kappa_{0}(-1)^m$ \cite{Marinica08,Bulgakov18,Hemmati19,HXu20,fengwu22}. Because $\kappa_{0}\ll\omega_{0}$ for low-Q leaky resonant modes, the resonant frequency of the Fabry-Perot BIC is close to $\omega_{0}$. Because the surface of each wavy grating monolayer is not flat, the effective distance between the two layers is slightly different from $d$. Within the range of wavelengths between 400 and 700 nm, each individual  monolayer wavy grating has two leaky resonant modes with resonant wavelengths of 492 and 411 nm, as shown in Fig. \ref{figure_1}. Consequently, Fabry-Perot BICs exist as $d$ equals 205.5$m$ nm or 246$m$ nm, with $m$ being an integer. The numerical results of the reflectance of the Fabry-Perot cavity for varying $d$ are plotted in Fig. \ref{figure_4}.

\begin{figure}[tbp]
\scalebox{0.6}{\includegraphics{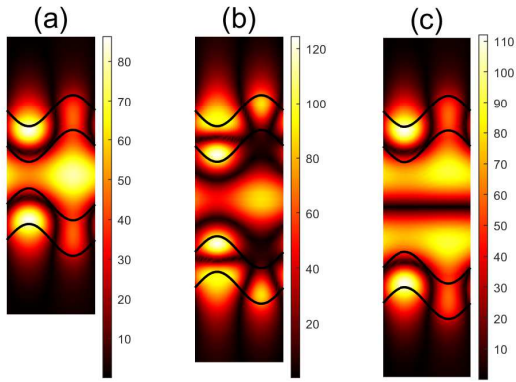}}
\caption{ The electric field distributions ($|E_{z}|$) at the peaks of one minus reflectance for $d=220$ nm at wavelength 493.335 nm in (a), for $d=400$ nm at wavelength 412.372 nm in (b), and  for $d=460$ nm at wavelength 490.152 nm in (c), with the amplitude of the incident plane being one.  }
\label{figure_5}
\end{figure}

The first Fabry-Perot BIC was generated from the leaky resonant mode of a monolayer wavy grating with a resonant wavelength of 492 nm. As $d$ increases from 160 to 300 nm, an anti-peak reflectance appears near 492 nm, with the center wavelength increasing. The width of the antipeak reached a minimum near $d=$220 nm. Thus, the resonant condition of the Fabry-Perot BIC is $d$ near 220 nm, which is less than half of 492 nm. The field pattern for the Fabry-Perot cavity with $d=$220 nm and the incident wavelength at the center of the anti-peak (i.e., 493.335 nm) is highly localized at the two layers of the wavy grating and spacing region, as shown in Fig. \ref{figure_5}(a). Within the spacing region, the field pattern does not have nodes, so that the standing wave has approximately the form $\cos(\pi y/d_{FP})$, where $y\in[-d_{FP}/2,d_{FP}/2]$ is the range of the spacing region, and $d_{FP}$ is the effective spacing distance. Because the field pattern within the spacing region has a larger amplitude at the x-coordinate with a larger interlayer distance, $d_{FP}$ is larger than $d$. For the Fabry-Perot BIC, $d_{FP}$ should be half of 492 nm so that $d$ is smaller than 246 nm. The same phenomenon occurs for the other Fabry-Perot BIC of the double-layer wavy grating. The node across the dielectric slab is caused by a standing wave along the axis of the dielectric slab with a wavy shape. The node along the wavy edge of the dielectric slab was due to the curvature of the slab. The total thickness of the structure that hosts the Fabry-Perot quasi-BIC mode is $d+2h+2w=610$ nm.

The second Fabry-Perot BIC was generated from the leaky resonant mode of a monolayer wavy grating with a resonant wavelength of 411 nm. When $d=400$ nm, the Fabry-Perot cavity approaches the resonant condition of the Fabry-Perot BIC. The field pattern at the center wavelength of the anti-peak (i.e., 412.372 nm) was also highly localized, as shown in Fig. \ref{figure_5}(b). Within the spacing region, the field pattern has two nodes, such that the standing wave has an approximate form of $\cos(2\pi y/d_{FP})$ with $y\in[-d_{FP}/2,d_{FP}/2]$. Within the dielectric slab, the field pattern had three nodes. The nodal line parallel to the x-axis is caused by the transverse standing wave of the second waveguide mode of the corresponding flat dielectric slab. The two nodes across the dielectric slab were caused by a standing wave along the axis of the dielectric slab.

The third Fabry-Perot BIC is the same as the first Fabry-Perot BIC, except that $d_{FP}$ is equal to the wavelength of the corresponding monolayer leaky resonant mode ( 492 nm). When $d=460$ nm, the Fabry-Perot cavity approaches the resonant condition of the Fabry-Perot BIC with a resonant wavelength of 490.152 nm. At the resonant wavelength, the field pattern had a node in the middle of the double layer, as shown in Fig. \ref{figure_5}(c). Thus, the standing wave within the spacing region has the form $\sin(2\pi y/d_{FP})$ with $y\in[-d_{FP}/2,d_{FP}/2]$. Similarly, another Fabry-Perot BIC is the same as the second Fabry-Perot BIC, except that $d_{FP}$ is equal to the half wavelength of the corresponding monolayer leaky resonant mode (i.e., 205.5 nm). However, this Fabry-Perot BIC is mixed with the third Fabry-Perot BIC, so it is not clearly exhibited by the reflectance.

\section{Conclusion}

The monolayer wavy grating hosts leaky resonant modes, whose resonant frequency and Q factor are tuned by the amplitude of the wavy shape $w$. In the band structure of the resonant modes, three BICs appear above the light cone and one resonant wavy mode with an infinite Q factor appears below the light cone. The double-layer wavy grating host Fabry-Perot BICs, which are due to the coupling between the leaky resonant modes of each layer. The condition of the Fabry-Perot BICs is that the effective spacing distance $d_{FP}$ between the two wavy gratings is equal to $m\lambda_{0}/2$, where $\lambda$ is the resonant wavelength of the leaky resonant mode of each individual monolayer and $m$ is an integer. Multiple-layer staggered wavy gratings in the far-field coupling regime exhibit high reflectance over a wide bandwidth. The total thickness of the structure that has high reflectance in wide bandwidth is no more than twice of the resonant wavelength. Because the Q factor of the Fabry-Perot cavity and the bandwidth of the high-reflectance can be controlled by the mechanical oscillation of the dielectric slab, enhancement of the optomechanical effect \cite{Aspelmeyer14} can be expected.

\begin{acknowledgments}
This project is supported by the Natural Science Foundation of Guangdong Province of China (Grant No.
2022A1515011578), the Project of Educational Commission of Guangdong Province of China (Grant No. 2021KTSCX064), the Startup Grant at Guangdong Polytechnic Normal University (Grant No. 2021SDKYA117), and the National Natural Science Foundation of China (Grant No.
11704419).
\end{acknowledgments}

\clearpage

\end{document}